\documentclass[sigconf]{acmart}
\AtBeginDocument{%
  }

\setcopyright{acmlicensed}
\copyrightyear{2018}
\acmYear{2018}
\acmDOI{XXXXXXX.XXXXXXX}
\acmConference[Conference acronym 'XX]{Make sure to enter the correct
  conference title from your rights confirmation email}{June 03--05,
  2018}{Woodstock, NY}
\acmISBN{978-1-4503-XXXX-X/2018/06}

\begin{document}

\title{Disrupting Cognitive Passivity: Rethinking AI-Assisted Data Literacy through Cognitive Alignment}

\author{Yongsu Ahn}
\affiliation{%
  \institution{Boston College}
  \city{Chestnut Hill}
  \state{Massachusetts}
  \country{USA}}
\email{anyon@bc.edu}

\author{Nam Wook Kim}
\affiliation{%
  \institution{Boston College}
  \city{Chestnut Hill}
  \state{Massachusetts}
  \country{USA}}
\email{nam.wook.kim@bc.edu}

\author{Benjamin Bach}
\affiliation{%
  \institution{Inria}
  \city{Bordeaux}
  \country{France}
}
\affiliation{%
  \institution{University of Edinburgh}
  \city{Edinburgh}
  \country{United Kingdom}
}
\email{benjamin.bach@inria.fr}


\begin{abstract}
  AI chatbots are increasingly stepping into roles as collaborators or teachers in analyzing, visualizing, and reasoning through data and domain problem. Yet, AI's default assistant mode with its comprehensive and one-off responses may undermine opportunities for practitioners to develop literacy through their own thinking, inducing \textit{cognitive passivity}. Drawing on evidence from empirical studies and theories, we argue that disrupting cognitive passivity necessitates a nuanced approach: rather than simply making AI promote deliberative thinking, there is a need for more dynamic and adaptive strategy through cognitive alignment---a framework that characterizes effective human-AI interaction as a function of alignment between users' cognitive demand and AI's interaction mode. In the framework, we provide the mapping between AI's interaction mode (transmissive or deliberative) and users' cognitive demand (receptive or deliberative), otherwise leading to either cognitive passivity or friction. We further discuss implications and offer open questions for future research on data literacy.
\end{abstract}



\received{20 February 2007}
\received[revised]{12 March 2009}
\received[accepted]{5 June 2009}

\maketitle

\section{Introduction}
Artificial intelligence is increasingly stepping into roles, once held by colleagues, mentors, and teachers, in answering questions about data analysis, critiquing visualization designs, and guiding analytical decisions. These feedback processes are often a cornerstone of improving data literacy through learning by doing in practice, ranging from analyzing, visualizing, to communicating domain insights in addressing real-world data problems \cite{alabood2023systematic,gray2013informal}.

However, studies have found that AI's default assistant mode (i.e., delivering comprehensive information in a transmissive, one-off response) can induce \textbf{cognitive passivity}, where practitioners receive answers rather than reason through problems, losing opportunities to enter into learning mode \cite{ahn2026answer, stadler2024cognitive}. This is particularly consequential for novices and developing practitioners, who benefit most from a scaffolded process that makes visible how to approach a data problem, what reasoning is needed at each stage, and why certain design choices serve communicative goals better than others \cite{collins1989cognitive}. Such a transmissive feedback mode can quietly erode the reasoning capacity that one may requires to foster data literacy \cite{gerlich2025ai}.

Then, how can we disrupt cognitive passivity? We find that it may involve a more nuanced approach than simply making AI promote deliberative, scaffolded interaction in all cases. Drawing on empirical evidence from prior studies of AI-assisted visualization feedback \cite{ahn2025understanding, kim2025tochi, ahn2026answer}, we propose that AI-mediated data literacy be understood as a function of alignment between users' cognitive demand and AI's interaction mode, what we referred to as \textbf{cognitive alignment} (Figure \ref{fig:figure}). A successful interaction over a data problem hinges not on simply whether AI scaffolds or transmits information, but on whether AI's mode matches what users' cognitive demand requires. When it does not, misalignment leads to either \textbf{cognitive passivity} (e.g.,  a novice is provided with a ready-made answer when their literacy is in need to reason through a design problem) or \textbf{cognitive friction} (e.g., an expert seeking quick confirmation is met with unnecessary probing questions).

While this paper aims to discuss the high-level notion for advancing AI-assisted literacy, we leave questions for the community to answer in the future research: (1) From the AI side, how can a system detect what a given moment demands, and adapt accordingly? (2) From the user side, how can learners develop the metacognitive awareness to recognize their own cognitive needs? These questions reveal that the challenge of AI-assisted data literacy is fundamentally an interaction design problem, not simply a matter of what content it should deliver. It requires understanding both how people develop data literacy and how they engage with AI systems. Overall, this paper proposes cognitive alignment as a lens and a set of open questions as directions for future interdisciplinary research on data literacy.

\section{Challenges: When AI meets Data Literacy}

\subsection{AI's default mode forces cognitive passivity}

A growing body of evidence suggests that AI's default, solution-oriented mode can reduce the cognitive engagement necessary for learning, a phenomenon termed differently across disciplines. Fan et al. \cite{fan2025beware} describe it as ``metacognitive laziness''---a reduction in self-regulatory behavior when AI provides ready-made answers. In scientific inquiry, Stadler et al. \cite{stadler2024cognitive} frame this as cognitive ease at a cost. These concerns are typically evidenced in outcomes: Bastani et al. \cite{bastani2024generative} found that students using GPT-4 improved practice performance by 48\% but scored 17\% worse on subsequent exams without AI access, and Kumar et al. \cite{kumar2025human} showed that while LLMs boosted individual creative output, they reduced the collective diversity of ideas generated.

In the specific context of data and visualization literacy, Bach et al. \cite{bach2023challenges} identify the risk of students over-relying on AI rather than engaging critically with visualizations as a key challenge for visualization education. Kim et al. \cite{kim2024vlhcc} found that students using ChatGPT in a visualization course appreciated its accessibility but gravitated toward accepting its suggestions without questioning design rationales. Kobiella et al. \cite{kobiella2024machine} document a broader consequence: young professionals reported reduced senses of accomplishment and self-efficacy when AI produced work comparable to their own. Ahn et al. \cite{ahn2026answer} pointed out a tendency for users to become dependent on AI suggestions without critically evaluating their appropriateness, especially when users lack sufficient domain knowledge to judge AI-generated recommendations.

\begin{figure*}[t]
\includegraphics[width=\textwidth]{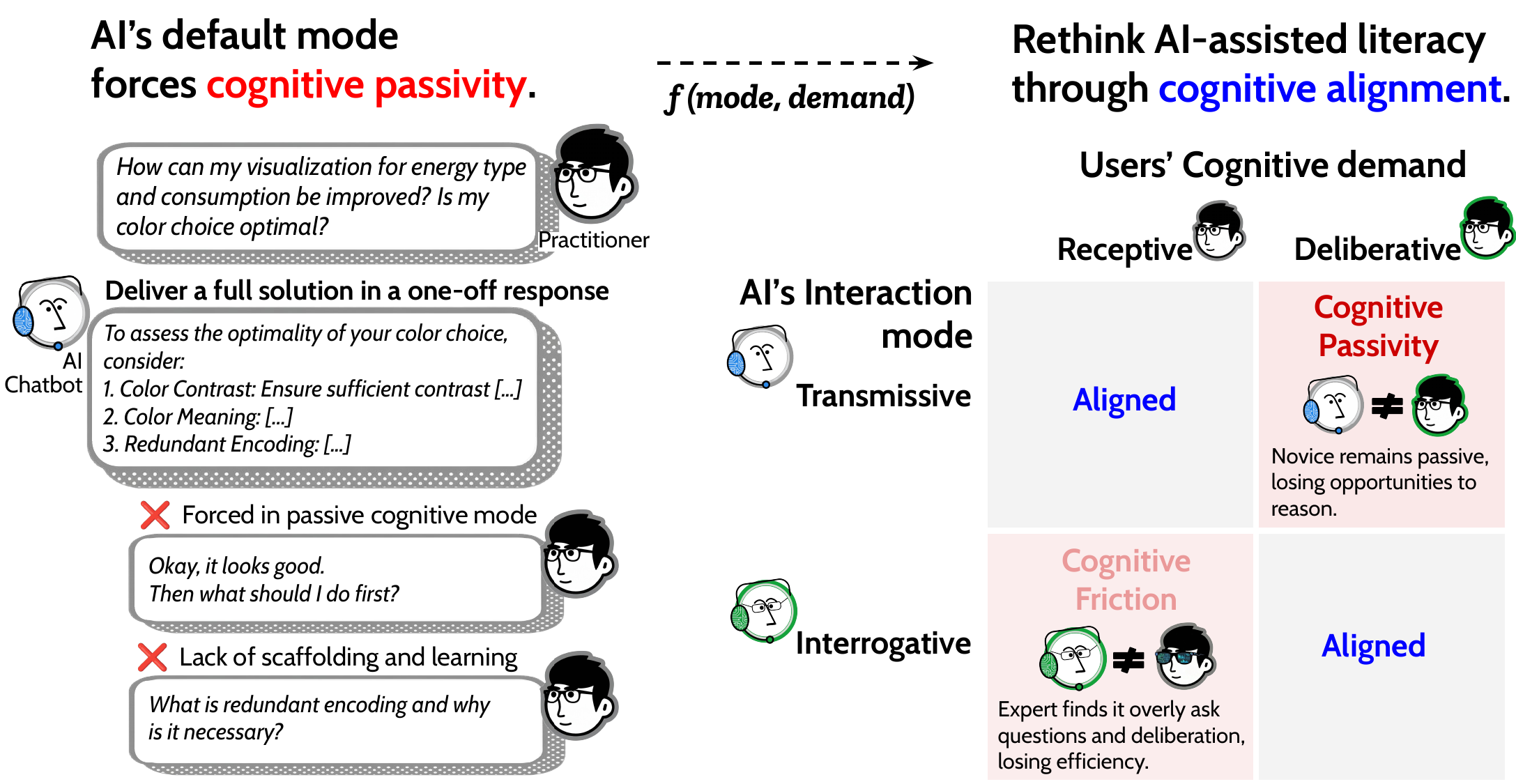}
\caption{\label{fig:figure} \textbf{Study overview. The current AI's default mode, producing ready-made responses at a one-off response,  easily induces cognitive passivity, harming opportunities for practitioners to foster data literacy. Beyond simply transforming AI into practicing reflective thinking, we suggest that this should be understood as a function of alignment between user's cognitive demand and AI's interaction mode. We provide a high-level description of potential cases where (mis)aligned cases can lead to a successful AI-assisted data analysis or different types of cognitive effects.}}
\end{figure*}

\subsection{AI short-circuits multi-level reasoning in data problem}

Fostering data literacy entails building a chain of reasoning across different levels of abstraction, concepts, and knowledge. Existing frameworks have characterized data literacy as a multi-level competency spanning from domain knowledge through analytical reasoning to representation and communication \cite{wolff2016creating, gummer2015building}. Addressing a data problem typically requires a sequence of decisions across these levels, such as understanding what question matters (domain), selecting an appropriate analytical approach (method), executing the analysis or design (technique), and interpreting what the result means and what it does not (critical evaluation).

However, in the AI-assisted data literacy context, studies have found that AI chatbots tend to systematically take a shortcut to this chain, predominantly responding at the execution level. Taking an example of visualization literacy as a well-formalized case, making a visualization decision, according to Munzner's nested model \cite{munzner2009nested}, consists of four different layers: domain problem characterization, data/task abstraction, visual encoding and interaction, and algorithm design. Ahn et al. \cite{ahn2026answer} empirically confirmed that, despite multi-level nature of a visualization problem, feedback from existing AI chatbot focused on encoding-level improvements, a low-level and visual element knowledge in a one-off response. It tended to overlook and skip underlying problems and notion in domain and problem-level definitions and issues, which are a cornerstone of how and what visual encodings are necessary to effectively communicate the insights through visual summaries.

\subsection{Disrupting cognitive passivity is possible, but context-dependent}

Drawing on cognitive theories and principles, studies have introduced interventions, such as cognitive forcing functions \cite{buccinca2021trust} and productive failure \cite{kapur2008productive}, to counteract passivity by adding obstacles or delays that prompt the learner to engage. Another line of work has restructured AI interactions through instructional scaffolding tailored to specific problem-solving contexts. Yan et al. \cite{yan2025effects} compared proactive GenAI agents that pose scaffolding questions against passive agents for visual learning analytics comprehension. Kazemitabaar et al. \cite{kazemitabaar2024codeaid} deployed CodeAid, an LLM-based programming assistant that generates pseudo-code and guided annotations rather than direct code solutions. In data visualization, an AI chatbot \cite{ahn2026answer} informed by the Cognitive Apprenticeship Model \cite{collins1989cognitive} was proposed to restructure the interaction to draw out reasoning through coaching, articulation prompts, and guided reflection rather than delivering evaluative feedback. Across these systems, the consistent finding is that interaction design determines AI's interaction mode: when the AI shifts from answering to guiding and scaffolding, learners engage more deliberatively.

However, disrupting passivity is not universally beneficial. Ahn et al. \cite{ahn2026answer} found that while reflective AI chatbot successfully shifted learners into deliberative engagement, these benefits were context-dependent. Participants in exploratory and developmental phases of their design process overwhelmingly preferred the scaffolded interaction, but those in evaluation phases preferred the baseline's directness. Moreover, experts often preferred simple and one-off responses to discuss a given issue, with less engagement in a scaffolded process due to their proficient knowledge. This suggests that cognitive passivity is bound to conditions, including users, tasks, and contexts, requiring to understand those boundaries for a more precise framework.

\section{Cognitive Alignment: Rethinking AI-assisted Literacy}

Given these evidence from previous work, we find that AI-assisted data literacy is not simply solved with a monotonic way of changing AI behaviors, but operates on a function of cognitive alignment between two aspects (Figure \ref{fig:figure}): 

\begin{itemize}
    \item \textbf{Cognitive demand} refers to a property of the learning moment---what kind of cognitive engagement the problem, task, and the learner's current literacy actually require. We distinguish two demands: \textit{receptive}, where the user and analysis context calls for simply taking in new information or confirming existing understanding, and 
    \textit{deliberative}, where the user and analysis context calls for reasoning through trade-offs and scaffolded processes from domain/problem understanding to problem solving.
    \item \textbf{Interaction mode} refers to a property of the interaction design 
    acting on the learner---what kind of cognitive engagement the AI's interaction elicits. We distinguish two modes: \textit{transmissive}, where AI delivers information, explanations, or solutions for the learner to receive, and \textit{interrogative}, where AI elicits reasoning through scaffolding, 
    questioning, or guided reflection.
\end{itemize}
 
The framing of cognitive mode is well-established across multiple literature such as  dual-process theory \cite{kahneman2011thinking}, ICAP framework \cite{chi2014icap}, desirable difficulties tradition \cite{bjork1994memory}, and automation complacency research \cite{parasuraman2010complacency}. While these literature have a solid theoretical background, our framework escalates it to a cognitive alignment problem in the human-AI interaction. Rather than simply making AI practice reflective process, we find that such an alignment determines how successfully practitioners solve the problem while not losing opportunities to grow and learn or failing to engage in AI systems.

Depending on how the alignment goes, based on the status of two dimensions, 
AI's interaction mode and practitioners' cognitive demand, there are four conditions with distinct implications 
for learning (Figure \ref{fig:figure}).

\begin{itemize}
    \item \textbf{Receptive demand, Interrogative mode} $\rightarrow$ 
    \textbf{Aligned.} When practitioners already possess relevant domain knowledge or have moved into a confirmatory stage of their analytical workflow, structured questioning and deliberative thinking in this context register as redundant rather than enriching the understanding. In such moments, the practitioner is not grappling with a design trade-off or reflecting on the design and analysis goals, but they are simply filling a discrete knowledge gap or responses straight to the inquiry. 
    \item \textbf{Deliberative demand, Interrogative mode} $\rightarrow$ 
    \textbf{Aligned.} When the learning moment requires reasoning, a scaffolded interaction that draws out the practitioner's own thinking is appropriate, engaging in structured support to work through revisiting fundamental problem understanding and analysis stages. This is where literacy develops, not from receiving the right answer, but from reasoning toward it.
    \item \textbf{Deliberative demand, Transmissive mode} $\rightarrow$ 
    \textbf{Cognitive passivity.} The learning context requires the practitioner to reason through a design trade-off, but the AI delivers a ready-made answer instead. The reasoning step---where literacy is built---is skipped. A novice who asks ``should I use a pie chart or a bar chart?'' receives ``use a bar chart'' without ever confronting a walk-through of how one serves the communicative goal better than the other. The practitioner may be given a better artifact but not better understanding.
    \item \textbf{Receptive demand, Interrogative mode} $\rightarrow$ 
    \textbf{Cognitive friction\footnote{The term \textit{cognitive friction} was originally introduced in the context of Human-Computer Interaction by \cite{cooper1999inmates}, referring to the resistance users feel when engaging with a complex system of rules. In this paper, we use the term to describe the discomfort users experience with AI, due to the misalignment between user and AI in their cognitive demand and interaction mode.}.} The practitioner needs a straightforward piece of information without unnecessarily scaffolding and revisiting the underlying processes, but the system responds with probing questions. It does not harm literacy but the practitioner may suffer from lack of efficiency and disengage from the system entirely.
\end{itemize}

\textbf{Demand often comes as opaque.} A critical complication is that cognitive demand---what cognitive mode one needs to engage in---does not always come as obvious to learners themselves. An experienced practitioner in the evaluation phase can recognize that they need confirmation, not deliberation, and may reasonably prefer a direct answer. Novices, on the other hand, are likely to blindly follow an easy path to get answers although they can benefit from a reflective learning process. In that case, the practitioner feels helped, the AI appears successful, but cognitive passivity becomes self-concealing---a blind spot induced by \textbf{demand opacity}, the degree to which the practitioner can recognize what cognitive mode fits into given context and benefits better. Several AI designs can facilitate this process, either adapt the mode of alignment based on user information and context, or set deliberative mode as default to minimize the cost of cognitive passivity, particularly when AI systems are served to learners and novices.

\section{Future Research Directions}
We discuss several directions for future research on advancing the AI-assisted data literacy through adaptive and interactive AI systems.

\subsection{Toward adaptive alignment in AI tools}
Overall, the framework argues that effective AI-assisted literacy requires interaction 
that is not uniformly scaffolded or uniformly transmissive, but dynamically aligned to what the moment demands. The type of cost may differ depending on expertise 
and context: practitioners either lose opportunities to reason (cognitive passivity) or lose the efficiency that sustains their willingness to engage (cognitive friction). In the adaptive process, cognitive demand is frequently opaque to learners themselves, 
particularly among novices who may default to seeking answers even when deliberation 
would serve them better. This poses some core design questions: can AI infer this from behavioral cues, such as question specificity, design stage, interaction history? When opacity is high, should AI systems default to deliberation despite the risk of frustrating users?

\subsection{Reasoning as a measure of data literacy}
In the era of AI-assisted data literacy, measuring literacy may need to shift from 
evaluating output quality to capturing reasoning: can learners explain their choices, apply principles to new contexts, and recognize when AI advice does not fit their situation? We pose a further question regarding users' metacognitive capacity: can learners tell when they have stopped reasoning and the AI has taken over? We find that the metacognition sits at the intersection of data literacy and AI literacy, prompting us to rethink what it means to be data literate when AI can produce polished outputs on a learner's behalf.

\subsection{Multi-level reasoning as design consideration}
AI consistently responds at the execution level, such as recommending a chart, running a test, flagging an anomaly, while skipping the reasoning on underlying problems and domains through which literacy develops. Can AI be designed to engage practitioners in understanding underlying problem and analysis tasks before delivering execution-level answers and one-off decisions, without burdening practitioners who do not need the guidance?

\subsection{Questions for the community}
Several questions extend beyond our evidence and worth contributions across disciplines:

\begin{itemize}
    \item When does passive consumption of AI guidance shift from useful scaffolding to harmful dependency?
    \item Can AI detect mode-demand misalignment in real time through behavioral cues?
    \item How does cognitive passivity manifest differently across contexts, such as visualization, statistics, data interpretation, AI-assisted analysis?
\end{itemize}

\section{Conclusion}

In this paper, we present that fostering AI-assisted data literacy depends on cognitive alignment, the match between an AI's interaction mode and a practitioner’s cognitive demand. Throughout the framework, we provide a lens for the community to rethink data literacy as a dynamic function of human-AI interaction and adaptive design. This highlights the need for future research on developing adaptive AI systems and modeling when and how to align with a user's shifting cognitive needs.


\bibliographystyle{ACM-Reference-Format}
\bibliography{main}


\end{document}